\documentclass[fleqn,usenatbib]{rasti}

\usepackage{newtxtext,newtxmath}

\usepackage[T1]{fontenc}

\DeclareRobustCommand{\VAN}[3]{#2}
\let\VANthebibliography\thebibliography
\def\thebibliography{\DeclareRobustCommand{\VAN}[3]{##3}\VANthebibliography}

\usepackage{graphicx}	
\usepackage{amsmath}	

\title[The nature of apparent transients on NGS--POSSI copy plates]{On the nature of apparent transient sources on the National Geographic Society--Palomar Observatory Sky Survey glass copy plates}

\author[N.~C.~Hambly \& A.~Blair]{
N.~C.~Hambly\thanks{E-mail: nch@roe.ac.uk}
and A.~Blair
\\
Institute for Astronomy, University of Edinburgh, Royal Observatory, Blackford Hill, Edinburgh EH9~3HJ, UK
}

\date{Accepted 2024 January 30. Received 2024 January 25; in original form 2023 October 30}

\pubyear{2024}

\begin{document}
\label{firstpage}
\pagerange{\pageref{firstpage}--\pageref{lastpage}}
\maketitle

\begin{abstract}

We examine critically recent claims for the presence of above--atmosphere optical transients in publicly--available digitised scans of Schmidt telescope photographic plate material derived from the National Geographic Society--Palomar Observatory Sky Survey. We employ the publicly available SuperCOSMOS Sky Survey catalogues to examine statistically the morphology of the sources. We develop a simple, objective and automated image classification scheme based on a random forest decision tree classifier. We find that the putative transients are likely to be spurious artefacts of the photographic emulsion. We suggest a possible cause of the appearance of these images as resulting from the copying procedure employed to disseminate glass copy survey atlas sets in the era before large--scale digitisation programmes.

\end{abstract}

\begin{keywords}
Instrumentation -- photography -- digitisation -- image classification -- machine learning
\end{keywords}



\section{Introduction}

The National Geographic Society--Palomar Observatory Sky Survey (hereafter POSSI) undertaken in the mid--twentieth century was the main deep and wide--angle optical reference of the sky for many decades following its publication 
\citep{1963bad..book..481M}. When subsequently supplemented with southern hemisphere counterpart surveys conducted with other Schmidt telescopes, notably the UK Schmidt Telescope at Siding Springs Observatory in Australia \citep{1984ASSL..110...25C}, an optical reference of the entire sky was created. Coupled with second--epoch surveys and the advent of large--scale digitisation programmes \citep{1995PASP..107..763L} whole--sky multi--colour and multi--epoch catalogues were created that became the cornerstone of survey astronomy prior to the advent of wide angle surveys with cameras employing large format digital detectors. The 1.2m Schmidt photographic surveys have been superseded in terms of depth and image quality by more recent programmes employing telescopes with larger apertures and sensitive~CCD cameras (e.g.~SDSS, \citealt{1998AJ....116.3040G}; {\it PanSTARRS}, \citealt{2016arXiv161205560C}). However it was not until relatively recently that a superior, deep {\it and truly all--sky} optical survey has been achieved with the advent of ESA--Gaia~\citep{2016A&A...595A...1G}.  Moreover the venerable digitised optical all--sky surveys remain relevant today as a snapshot of the sky at epochs during the latter half of the last century. They are of course particularly important as an early epoch reference for current and future time--domain astronomical surveys. 

Recently a number of articles have appeared in the astronomical literature noting the presence of a significant population of apparently transient sources on POSSI `E' (red) plate scans \citep[][and citations thereof]{2021NatSR..1112794V}. The data analysed came from both the Digitised Sky Survey \citep[DSS;][]{1992ASSL..174...87L} and the SuperCOSMOS Sky Survey  \cite[SSS;][]{2001MNRAS.326.1315H,2001MNRAS.326.1295H,2001MNRAS.326.1279H}. While the original discovery paper discusses various plausible non--astronomical origins for the~9 transients identified, subsequent work \citep[e.g.][]{2022MNRAS.515.1380S, 2024MNRAS.527.6312S} is firmly based on the premise that these apparent transients represent real, above--atmosphere astronomical phenomena. If real, causality arguments in conjunction with the appearance and disappearance, over angular scales of tens of arcminutes, of multiple simultaneous detections in 45~minute exposures separated by 6~days require the sources to be well inside the solar system. \cite{2022arXiv220406091V} propose a population of highly reflective, glinting objects in near--Earth orbit as the source of the transients. No such population has been noted in more recent, deeper surveys such as SDSS and {\it PanSTARRS}, but if it truly exists there would be major implications for current and future deep, high time--cadence surveys, e.g.~the Vera Rubin Observatory Legacy Survey of Space And Time \citep[LSST;][and references therein]{2023ApJS..266...22S}. Such surveys are already facing significant contamination problems from constellations of artificial satellites currently being launched into low Earth orbit \citep[][not that this was a concern at the mid twentieth century epoch of POSSI of course]{2022ApJ...941L..15H}.

In this paper we present an independent analysis of the source photographic plate material and SSS scans used in passing by \cite{2021NatSR..1112794V}. We argue against the premise that these are real above--atmosphere transients and suggest an alternative, mundane explanation as to their origin.

\section{Observations, methods and results}

The 1.2~m Palomar Oschin and UK Schmidt telescopes employed $355\times355$~mm$^2$, 1~mm thin glass plates in a vacuum--clamping plate holder for optimum focus over the focal surface of the imaging optics. The delicate survey originals were then copied on paper (in the case of POSSI), film and glass for distribution to observatories and research institutes for the purposes of inspection and quantitative densitometry \citep{1992ASSL..174...11M}. It was primarily the glass copies that were digitised at scale by various scanning programmes in the latter half of the twentieth century \citep[][and references therein]{2001MNRAS.326.1279H}. It is our understanding that of the major all--sky digitisation programmes undertaken, only that of the USNO \citep{2003AJ....125..984M} had access to the POSSI originals for scanning. The SSS (Hambly et al. 2000a) and DSS (B.~McLean, personal communication; see also \citealt{1995PASP..107..763L}) employed glass copies of POSSI. 
The primary method for atlas copy production was contact printing \citep{1980AASPB..23....3W}. In this process intermediate glass positives, and then a second--generation copy negative from that positive, were made for each original. Exposures were made in vacuum and using UV light for optimum focus and resolution over the full $6\times6$~deg$^2$ field of view. Hence the DSS and SSS data used in the study of \cite{2021NatSR..1112794V} are in fact derived from independent machine scans and software processing of independent glass copies, those independent copy negatives having been generated from the same copy positive, itself a copy of the one original POSSI negative. 

\subsection{Visual inspection of copy plates}

\begin{figure*}
	\centerline{\includegraphics[scale=0.35]{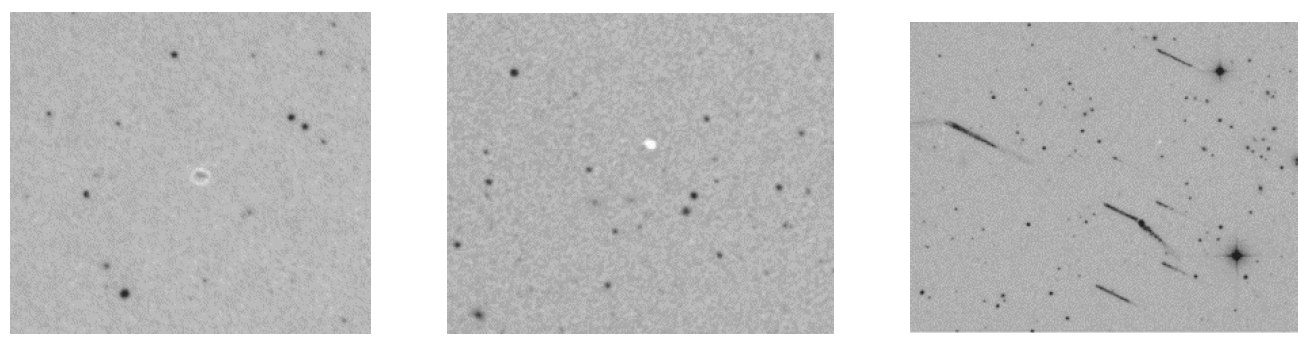}}
    \caption{Examples of imaging defects on glass copy plates of POSSI fields E0070 and E0086 in the atlas set held at the Royal Observatory Edinburgh. On the left is what appears to be a bubble in the thin emulsion coating on the plate; in the middle is an emulsion hole; and on the right are examples of dark marks that could plausibly be produced via emulsion scratches on the intermediate copy positive.}
    \label{fig:blemishes}
\end{figure*}

The copy set scanned during the SSS programme is to this day housed in the plate archive of the Royal Observatory, Edinburgh and is available for close inspection work. We inspected the locally held glass copy plates of POSSI survey fields E0070 and E0086 visually to verify the presence of the transients identified on the former in the publicly available digital scans by \cite{2021NatSR..1112794V}. We confirmed the presence of~9 point--like detections on E0070 at the positions reported, albeit appearing (to the eye aided by microscope) as generally rather more concentrated and circular than typical stellar images of the same estimated brightness. We noted, also as originally reported, that they are absent in the overlapping plate E0086 (exposed 6~days later) but furthermore that there are similar detections present on E0086 that are not present on E0070. At the same time we noted also the presence of various blemishes on the POSSI copy plates and we present some examples in Fig.~\ref{fig:blemishes}. The presence of emulsion holes is particularly germane (see later).

\subsection{Image profile statistics}

\begin{figure}
	\centerline{\includegraphics[scale=0.34]{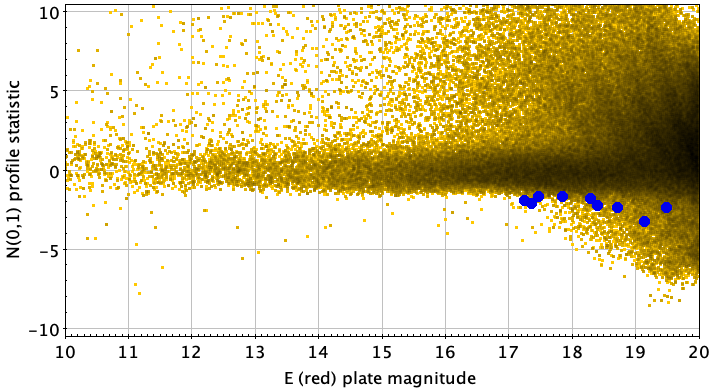}}
    \caption{Distribution of the stellar profile statistic $\eta$ for the SSA catalogue detections in POSSI survey field E0070 versus red magnitude~$E$ in the range $10 < E < 20$ and $-10.0 < \eta < +10.0$. The central $\mathcal{N}(0,1)$ distribution at any magnitude is occupied by point sources; galaxies and blends scatter to positive values of the statistic while sharp, noise features tend to scatter to negative values. Blue circles highlight the data points of the transient sources identified by Villarroel et al.~(2021).}
    \label{fig:profilestat}
\end{figure}

Given the appearance of the blobs as somewhat sharper than is typical for stars we extracted profile information from the SSS catalogues via the SuperCOSMOS Science Archive \citep[SSA;][]{2004ASPC..314..137H}. Amongst a set of image parameters measured in the digitised plate data (position, brightness, morphology) this provides a {\it profile statistic} (hereafter denoted $\eta$) for each detected source \citep{2001MNRAS.326.1295H}. This is a continuously distributed, zero--mean unit--weight statistic distilled from 1d radial profile information. That information is derived from the `areal profile' as determined at re-thresholded intensity levels above the threshold isophote as part of the image parameter analysis stage using the algorithm of \cite{1983ESASP.201..195B}\footnote{available online at \url{https://articles.adsabs.harvard.edu/pdf/1983ESASP.201..195B}}. The normalisation is done with respect to the dominant population of point--like sources on a given plate. The resulting $\mathcal{N}(0,1)$ statistic quantifies how point--like each detection is compared to the typical stellar profile for the plate and, amongst other benefits, provides the means for discrete image classification cuts. For example in the SSA stellar images have $-3 < \eta < +2.5$; galaxies with shallower profiles have $\eta > 2.5$; and spurious noise images with generally sharper (than stellar) profiles have $\eta < -3.0$. Fig.~\ref{fig:profilestat} shows the distribution of $\eta$ versus magnitude for detections in the SSA from POSSI plate E0070 (SSA {\tt plateId} = 327750; see example queries in the Appendix). Also highlighted are the~9 potential transients identified by \citet{2021NatSR..1112794V} and it is noteworthy that all lie on the negative side of the $\mathcal{N}(0,1)$ distribution in the range $-3 \lessapprox \eta < -1$. While these result in a rough discrete classification flag of 2 ($\equiv$~star) in the SSA for each individual detection in~8 of the ~9 cases (one having $\eta = -3.2$ is classed as noise), the combined probability of drawing nine values having $\eta \lessapprox -1.0$ from a truly normal distribution is~$\mathcal{O}(0.1^9)$, i.e.~vanishingly small.

\subsection{Image classification via Machine Learning}

In order to put the analysis of image morphology on a more objective statistical footing we undertook a simple machine learning (ML) image classification exercise. We selected a subset of the morphological parameters available from the SSA and trained up a random forest decision tree classifier in three classes (star, galaxy, spurious) using independent modern sky survey catalogues as a source of highly reliable training data. The procedure undertaken is outlined in the following sections (where we have denoted public database table and column names in mono-type for convenience and clarity). 

\subsubsection{Plate catalogue data selection}

We created individual plate catalogue selections from the SSA for fields~E0070 and~E0086 by applying cuts to retain isolated, well--exposed detections with no
obvious quality issues: {\tt blend = 0}; {\tt sMag}~$<19.5$; {\tt quality}~$< 128$.
We retained the following~6 catalogue parameters as features on which to train and subsequently classify: profile statistic~$\eta$ {\tt prfStat}; photographic plate magnitude {\tt sMag}; image area {\tt area}; intensity weighted image location on plate {\tt xCen, yCen}; and eccentricity~$e$ computed from the intensity weighted semi--major and semi--minor ellipse fit parameters {\tt aI} and {\tt bI} (an example SSA database query is provided in the Appendix).

\subsubsection{Training data: high reliability stars}
\label{sec:reliablestars}

A selection of highly reliable, isolated stars covering each field was created using {\it Gaia}~DR3 \citep{2023A&A...674A...1G}. We made an astrometric reliability cut following \cite{2021A&A...649A...4L} via the `renormalised unit--weight error' statistic ({\tt ruwe}~$<1.4$; again, an example query is provided in the Appendix). This selection was then proper motion corrected to the observation epoch of the plate and paired with the plate catalogue (data as defined above) using a proximity criterion of 1~arcsec, this being dictated by the likely uncertainties in the SSS plate astrometry (Hambly et al. 2001c), to create a list of high reliability stars detected on the plate.

\subsubsection{Training data: high reliability galaxies}

A selection of highly reliable galaxies covering each field was created using PanSTARRS PS1--DR2 \citep{2020ApJS..251....7F}. Tables {\tt ObjectThin} and {\tt StackObjectThin} were joined to a provide multiple--detection, multi--band catalogue including PSF and Kron magnitudes and source flags. Our selection required detection in both PS1~$r$ and~$i$ and we applied the recommended star--galaxy separation criterion {\tt iPsfMag}$-${\tt iKronMag}$>$0.05 \citep{2014MNRAS.437..748F}. A brightness cut was applied as {\tt rPsfMag$<$21}.
Once again the training input data set was created by proximity pairing the high reliability catalogue of galaxies to the plate catalogue using a proximity criterion of 1~arcsec.

\subsubsection{Training data: high probability spurious detections}
\label{sec:spurious}

Our approach to defining a training set of high likelihood spurious plate detections was to use highly complete star and galaxy catalogues as defined above but without the quality criteria, and negate the pair association with a relaxed proximity criterion. We defined a plate catalogue entry as likely spurious if there was no associated Gaia stellar or PS1 galaxy entry of any kind within 5~arcsec of its measured position. In this way we ensured that if there was any chance of a plate catalogue entry being real it would not be included in the spurious detection training data. Needless to say the transient catalogue detections on E0070 were excluded from the training set.

\subsubsection{Training and testing}

Following standard practice in defining ML training data the star and galaxy sets were sub-sampled down to be the same size as the spurious detection catalogue (this being the smallest of the three) and then all three were split 80\%/20\% to give independent training and validation sets numbering~$\approx 10,000$ and~$2,500$ entries respectively. We employed the random forest implementation in scikit--learn \citep{JMLR:v12:pedregosa11a} for our decision tree classifier. 

\begin{figure*}
	\centerline{\includegraphics[scale=0.23]{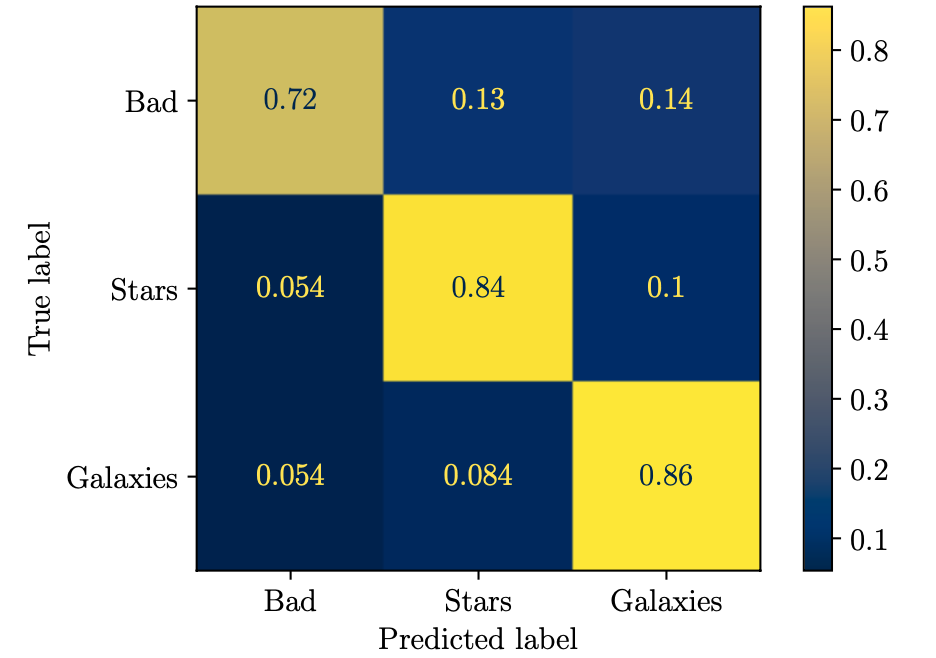}
	            \includegraphics[scale=0.3]{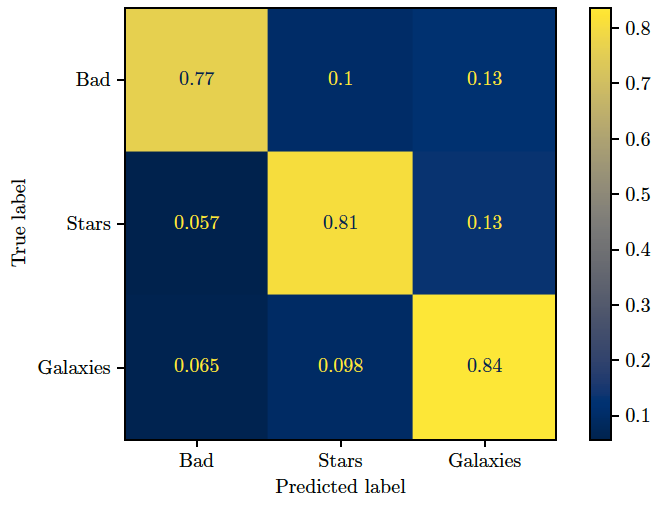}}
    \caption{Normalised confusion matrix amongst the true/false positive/negatives when training for 3 detection classes in the data sets for field E0070 (left) and E0086 (right). True positives are on the diagonal from upper left to lower right in each case. Field--to--field differences indicate likely random uncertainties in the procedure; the magnitude cut of E~$<19.5$~mag avoids the regime where emulsion noise introduces the strongest dependencies on signal--to--noise ratio in the results.}
    \label{fig:confusion}
\end{figure*}

Automated hyper-parameter tuning \citep{JMLR:v20:18-444} was achieved using the scikit--learn HalvingRandomSearchCV module. Two key hyper-parameters in random forest decision trees are the number of trees (also known as the number of estimators) and the number of features (up to a maximum of all those available) that are employed at each decision node \citep{efron1994introduction, Hastie2009}.
Such hyper-parameters influence execution speed, reliability and need to be chosen with care to avoid over--fitting. We examined the
`out--of--bag' (OOB) error for our training sets and found an optimum configuration (minimum number of estimators beyond which there is no significant improvement in the OOB error) of $\approx 100$~trees with a restriction on the maximum number of features used as square--root (E0070) or base--2 logarithm (E0086) of the number available, i.e.~$\surd 6$ or $\log_26$. The resulting scikit--learn hyper-parameters for training independent random forest classifiers for the two adjacent fields~E0070 and~E0086 are given in Table~\ref{tab:hyperparams} and in Figure~\ref{fig:confusion} we show the confusion matrix for each field. Table~\ref{tab:precision} quantifies precision (= true positives divided by sum of true and false positives), recall (= true positives divided by sum of true positives and false negatives) and F1 score \citep[= weighted mean of precision and recall,][]{FAWCETT2006861}. Finally in Table~\ref{tab:importance} we show the feature importance output as part of the training. As expected it is the profile statistic that carries most weight in classifying the images while position on the plate is relatively unimportant.

\begin{table}
	\centering
	\caption{Random forest hyper-parameters for the two fields.}
	\label{tab:hyperparams}
	\begin{tabular}{lcc} 
		\hline
		Hyper-parameter & \multicolumn{2}{c}{Value in field}\\
                        & E0070 & E0086 \\
		\hline
		{\tt max\_depth} & 22 & 3 \\
		{\tt min\_samples\_split} & 23 & 6 \\
		{\tt min\_samples\_leaf} & 11 & 7 \\
        {\tt max\_features} & $\surd6$ & $\log_2(6)$\\
        Number of trees & 128 & 128 \\
		\hline
	\end{tabular}
\end{table}

\begin{table*}
	\centering
	\caption{Precision of random forest classifiers trained and optimised as described in the text for the two adjacent POSSI fields. The F1 score is a weighted mean of
    precision and recall (Fawcett~2006).}
	\label{tab:precision}
	\begin{tabular}{lcccccc} 
		\hline
		Class & \multicolumn{6}{c}{Metric}\\
                        & \multicolumn{3}{c}{---~E0070~---} & \multicolumn{3}{c}{---~E0086~---} \\
            & Precision & Recall & F1 score & Precision & Recall & F1 score \\
		\hline
		Star   & 0.80 & 0.84 & 0.82 & 0.80 & 0.81 & 0.81 \\
		Galaxy & 0.78 & 0.86 & 0.82 & 0.86 & 0.84 & 0.81 \\
		Bad    & 0.87 & 0.72 & 0.79 & 0.76 & 0.77 & 0.80 \\
		\hline
	\end{tabular}
\end{table*}

\begin{table}
	\centering
	\caption{Decision tree feature importance for the random forest classifiers in the two adjacent fields.}
	\label{tab:importance}
	\begin{tabular}{lrr} 
		\hline
		Feature & \multicolumn{2}{c}{Relative importance}\\
                & \multicolumn{2}{c}{in field } \\
                        & \multicolumn{1}{r}{E0070} & \multicolumn{1}{r}{E0086} \\
		\hline
		{\tt prfStat} & 50.5\% & 45.8\% \\
		{\tt area}    & 16.2\% & 17.2\% \\
		{\tt e}       & 16.6\% & 16.9\% \\
        {\tt sMag}    & 10.6\% & 12.7\% \\
        {\tt xCen}    &  3.3\% &  3.9\% \\
        {\tt yCen}    &  2.8\% &  3.5\% \\
		\hline
	\end{tabular}
\end{table}

\subsubsection{Classifying the whole plate detection catalogues}

\begin{figure}
	\centerline{\includegraphics[scale=0.22]{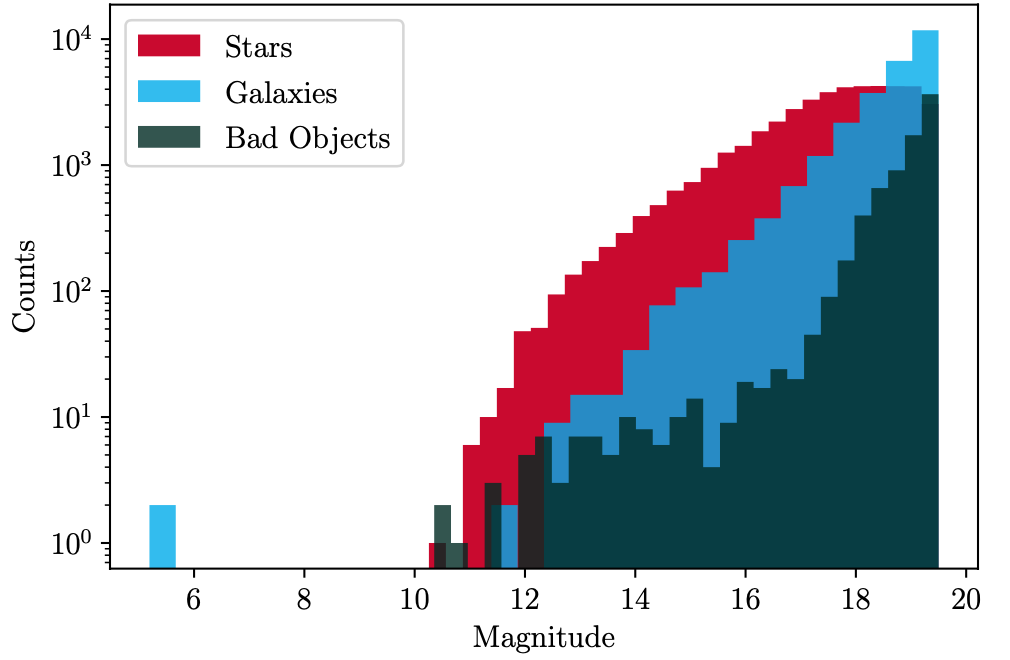}}
    \caption{Number--magnitude counts in the three discrete classes for field~E0070 as determined via the ML classifier as described in the text.}
    \label{fig:classhist}
\end{figure}

The individual trained classifiers described above were applied to the full SSA catalogues for the adjacent fields~E0070 and~E0086. For the former this resulted in $\approx 45,000$~stellar, $\approx27,000$ galaxy, and $\approx8000$ bad detections brighter than E~=~19.5 mag. A histogram is shown in Figure~\ref{fig:classhist} and exhibits logarithmic number--magnitude counts for stars and galaxies that conform to expectations, with shallower star counts and steeper, log--linear galaxy counts that cross at around 18th magnitude. The number counts of sources classified as bad rise dramatically towards the faint end consistent with the presence of a population of faint detections in the emulsion noise, originating from optical imaging artefacts chopped up by the automated image analysis software, or from foreign body contamination (e.g.~dust and hair etc.) on the scanned plates \citep[see e.g.][]{2004MNRAS.347...36S}.

\begin{figure}
	\centerline{\includegraphics[scale=0.28]{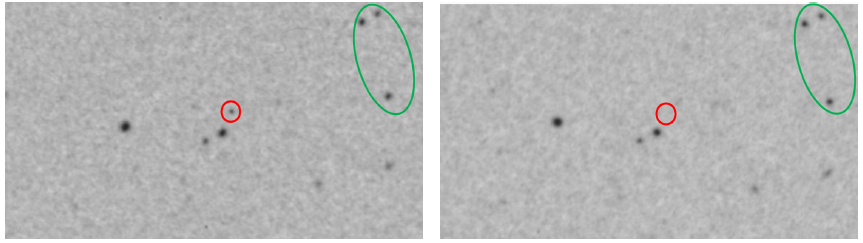}}
	\centerline{\includegraphics[scale=0.28]{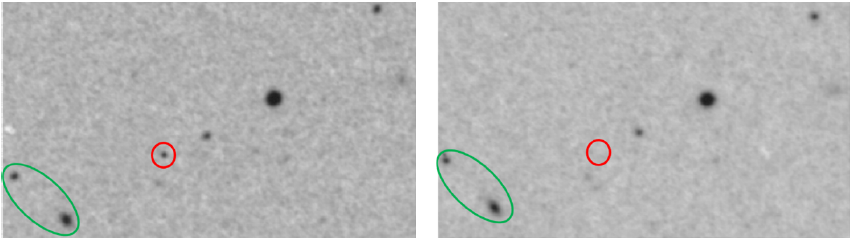}}
    \caption{Examples of two bad (according to our ML classifier) detections masquerading as transient sources in the overlap regions of POSSI fields E0086 (right--hand panels) and~E0070 (left--hand panels), being present in the former while absent in the latter. Several real astronomical sources common between the two fields in each case are circumscribed in green while the bad detection is circled in red.}
    \label{fig:moretransients}
\end{figure}

As another method of validating our results, albeit in a necessary--but--insufficient sense, we employed the small overlap region between the two adjacent fields. We took the lists of detections classified as bad in the two fields and searched for positional coincidences within a tolerance of 1~arcsec, and there were none as we would expect. Furthermore, the availability of the overlap region enabled us to search for apparent transients appearing on E0086 that are not present on~E0070, i.e.~the opposite situation to that described in \cite{2021NatSR..1112794V}. We found more than~100 detections on E0086 in the overlap region with brightness, eccentricity and area similar to the~9 detections on~E0070 and some examples are shown in Figure~\ref{fig:moretransients}. Of course, there will be a $\approx10$\% level of false, contrary classifications of stars as galaxies and galaxies as stars in the overlap region between the two plates, as indicated by the confusion matrices in Figure~\ref{fig:confusion}.

From the application of our trained classifier for field~E0070 we note that all~9 of the transients identified in \cite{2021NatSR..1112794V} are classed as bad with probabilities amongst the various classes as shown in Table~\ref{tab:ml6results}.

\begin{table}
	\centering
	\caption{ML classifier results for the~9 apparent transients of Villarroel et al.~(2021) in terms of percentage relative probability amongst the three classes.}
	\label{tab:ml6results}
	\begin{tabular}{cccc} 
		\hline
		Transient & \multicolumn{3}{c}{Class probability \%}\\
           no.     & bad & star & galaxy\\
		\hline
1 & 80.4 & 19.6 &  0.0 \\
2 & 99.3 &  0.7 &  0.0 \\
3 & 95.9 &  4.0 &  0.1 \\
4 & 62.4 & 37.1 &  0.5 \\
5 & 98.3 &  1.2 &  0.5 \\ 
6 & 74.5 & 24.6 &  0.9 \\
7 & 67.3 & 32.5 &  0.2 \\
8 & 81.4 & 18.4 &  0.2 \\ 
9 & 85.7 & 14.0 &  0.3 \\
		\hline
	\end{tabular}
\end{table}

\section{Discussion}

The precision of the ML classifier as described above is~81\% for field E0070. This is usable for the present purpose, at least in the sense that of the nine detections under investigation we might expect around~8 to be correctly classified, so the fact that one or two cases in Table~\ref{tab:ml6results} have relatively lower distinction between the bad and stellar classes should not be too concerning. In any case, none have any significant probability of being galaxies which is of course not surprising given their sharp profiles as indicated by their significantly negative profile statistics. The relatively low precision does not compare particularly favourably with previous applications of ML techniques to image classification of parameterised images on photographic plates. For example \cite{1992AJ....103..318O}, \cite{1995AJ....109.2401W} and \cite{Odewahn_2004} applied ML techniques to catalogues derived from photographic plates and achieved significantly higher precision. However in these works only two classes (star and galaxy) were considered whereas in this study we have a third class whose occupation of the low--dimensionality feature parameter space may well be less distinct than that of stars and galaxies. Moreover some reliability tests in those earlier works were done internally in overlap regions since at that time deep, wide--angle and high angular resolution catalogues from missions such as {\it Gaia} and {\it PanSTARRS} were not available for training and validation. Furthermore later work  employed data from measures of second--generation POSS original plates employing fine--grained Kodak IIIa emulsions whereas the data analysed here are from copy plates employing older, coarser 103a emulsions (as were state--of--the--art at the time of POSSI). On the subject of modern (i.e.~late twentieth--century) hyper-sensitised, fine grained Kodak emulsions it is interesting to note that there is a large literature on the appearance of spurious microdots \citep[see e.g.~][and references therein]{1988astr.conf...28G} but no such artefacts have been reported as occurring on the older coarse--grained varieties.

Neither \cite{2021NatSR..1112794V} nor any of the follow--up works discuss the provenance of the photographic plate material scanned in creating the images on which their analyses are based yet the reproduction of glass copy plates from the survey originals is clearly an important consideration. Emulsion flaws will be present on original negatives, contact positives derived from them as the first step in creating copies, and any paper, film or glass copy negatives derived from those positives. While the presence of the transients in independent scans (DSS and SSS) does indeed eliminate the respective digital scanning procedures, hardware and software as introducing the images under consideration here, it does not follow that those images must be present on the original negatives and are astronomical in origin. For example, suppose that there were a significant number of bubbles and holes (e.g.~Figure~\ref{fig:blemishes}) in the emulsion coatings of the plates used as positives in the production of the glass copy negatives. These would result in the appearance of dark spots on those copies since they would appear as low density spots in the intermediate positives in the same manner as real astronomical sources. This would easily explain the presence of the~9 apparent transients and many hundreds or even thousands more on each POSSI copy plate, i.e.~it is likely that the entire survey is peppered with such isolated detections. Unfortunately we cannot make any quantitative assessment of the prevalence of holes and bubbles like those illustrated in Figure~\ref{fig:blemishes} since the automated analysis in the digitised sky survey programmes considered only dark detections in an otherwise light background as potential sources and densities below that of the sky were ignored.

Clearly it would be most instructive to analyse (or at the very least, visually inspect) the original POSSI negative plates and the intermediate copies. The originals are, we understand, archived in the Carnegie Observatories plate vault as part of the plate archive holdings of the Observatories of the Carnegie Institution for Science\footnote{\url{https://sites.astro.caltech.edu/palomar/media/archival.html\#science}}. The existence and, assuming they were retained for posterity, location of the intermediate positives used in copy atlas production is unknown to us.
As noted previously the only large--scale digitisation and catalogue generation programme to scan original glass negatives of POSSI was that of the USNO \citep{2003AJ....125..984M}. Unfortunately the image data from those scans appears to be no longer available and furthermore USNO--B1.0 catalogue entries were created only for detections paired between at least two plates of different epoch or colour. Detections appearing on one plate only (such as those under consideration here) were never retained in USNO--B1.0 catalogue anyway, even supposing they were present on the original negative plates.

\section{Conclusions}


We have undertaken an independent study of~9 apparent optical transients identified by \cite{2021NatSR..1112794V} in publicly available digitised scans of POSSI copy plates. We have verified the presence of those detections on locally held copy plates from which SSS scans were made, noting subtle differences between them and other stellar images of similar brightness. We also noted the presence of emulsion blemishes (holes) during our visual inspection. We have made an objective, quantitative and statistical analysis of the morphological properties of the publicly available catalogues of all detections on the same plates, derived during the production of the whole--sky SSS, from two adjacent fields. We find that a) the image profiles of the transients are significantly sharper than typical stellar images on the plates; b) that an ML decision--tree classifier badges the images as spurious with high probability; c) that similar examples of apparent transients are present on the copy plate of the adjacent field; and finally d) that there are many hundreds of similar images on both plates in the overlap region between the two fields. We suggest one likely mechanism for the origin of at least some of these apparent transients as being emulsion holes on the intermediate positive plates used during reproduction of the copy sets. We therefore caution that digitised all--sky survey catalogues derived from the POSSI glass copies are likely peppered with these isolated false detections and that great care must be exercised when interpreting the publicly available digitised images or when making samples of unpaired catalogue records derived from them.

\section*{Acknowledgements}


We thank two anonymous referees for their positive reception and constructive comments on the original version of this paper. This research has made use of data obtained from the SuperCOSMOS Science Archive,
prepared and hosted by the Wide Field Astronomy Unit, Institute for Astronomy, University
of Edinburgh, which is funded by the UK Science and Technology Facilities Council.
This work has made use of data from the European Space Agency (ESA) mission Gaia
(https://www.cosmos.esa.int/gaia), processed by the Gaia Data Processing and Analysis
Consortium (DPAC, https://www.cosmos.esa.int/web/gaia/dpac/consortium). Funding
for the DPAC has been provided by national institutions, in particular the institutions
participating in the Gaia Multilateral Agreement.
The Pan-STARRS1 Surveys (PS1) and the PS1 public science archive have been made possible
through contributions by the Institute for Astronomy, the University of Hawaii, the
Pan-STARRS Project Office, the Max-Planck Society and its participating institutes, the
Max Planck Institute for Astronomy, Heidelberg and the Max Planck Institute for Extraterrestrial
Physics, Garching, The Johns Hopkins University, Durham University, the University
of Edinburgh, the Queen’s University Belfast, the Harvard-Smithsonian Center for Astrophysics,
the Las Cumbres Observatory Global Telescope Network Incorporated, the National
Central University of Taiwan, the Space Telescope Science Institute, the National Aeronautics
and Space Administration under Grant No. NNX08AR22G issued through the Planetary
Science Division of the NASA Science Mission Directorate, the National Science Foundation
Grant No. AST-1238877, the University of Maryland, Eotvos Lorand University (ELTE),
the Los Alamos National Laboratory, and the Gordon and Betty Moore Foundation.

\section*{Data Availability}

All data used in this study are publicly available from on--line archive systems such as the SuperCOSMOS Science Archive~\citep{2004ASPC..314..137H}, the Gaia Archive~\citep{2020ASPC..522..307S} and the Barbara A.~Mikulski Archive for Space Telescopes~\citep[MAST;][]{mast2021}. Some example queries are provided for convenience in the Appendix.



\bibliographystyle{myrasti}
\bibliography{refs} 




\appendix

\section*{APPENDIX: Example archive queries for data sets employed in the study}


Plate details and internal scanning housekeeping for the POSSI--E plates studied herein (E0070 and E0086) can be derived with the following query (cut--and--paste into the `Freeform SQL' web form\footnote{\url{http://ssa.roe.ac.uk/sql.html}} in the SSA having first selected `Full SSA' as the database):
\begin{verbatim}
SELECT * 
FROM Plate 
WHERE plateNum IN (70,86) AND 
      emulsion LIKE "103a" AND 
      filterID LIKE "E"
\end{verbatim}
This shows the internal plate identifiers for the relevant catalogue subsets (plate E0070 has identifier 327750 and E0086 is 327766). For example, the data plotted in Figure~\ref{fig:profilestat} can be generated as follows:
\begin{verbatim}
SELECT sMag, prfStat
FROM   Detection
WHERE  plateId = 327750 AND sMag < 20.0 AND
       ra BETWEEN 205.0 AND 220.0 AND 
       dec BETWEEN +26.0 and +34.0
\end{verbatim}
where the final position predicate is a speed optimisation trick to force an indexed search of the $\sim$10~billion row detection table. As another example, internal detection object identifiers for the~9 transient sources identified by positions quoted in~\cite{2021NatSR..1112794V} or measured from an SSS (in the case of the~3 detections not having quoted positions) can be derived using the SSA cone--search facility\footnote{\url{http://ssa.roe.ac.uk/radial.html}} and then included in a query to separately identify their data for over-plotting in Figure~\ref{fig:profilestat}:
\begin{verbatim}
SELECT sMag, prfStat, ra, dec
FROM   Detection
WHERE  objId IN 
    (1409569612299549, 1409569612298291, 
    1409569612298315, 1409569612295478, 
    1409569612295920, 1409569612296943,
    1409569612291853, 1409569612291887, 
    1409569612291586)
\end{verbatim}
A complete SSA catalogue containing a column projection including only those features used in the ML part of the analysis for one or other plate can be generated as follows (column explanations are provided in the database schema browser\footnote{\url{http://ssa.roe.ac.uk/www/ssa\_browser.html}} for all tables):
\begin{verbatim}
SELECT DISTINCT ra,dec, aI, bI, class, blend, 
       quality, prfStat, sMag, 
       sqrt(1.0 - (bI*bI)/(aI*aI)) AS e
FROM Detection
WHERE plateId = 327750 AND surveyId = 5 AND
    ra BETWEEN 205.0 AND 220.0 AND 
    dec BETWEEN +26.0 AND +34.0 AND
    blend = 0 AND quality < 128 AND
    sMag < 19.5
\end{verbatim}
(note that owing to the construction of the SSA database there are duplicate catalogue records for POSSI--E plate detections and these are filtered out above using the DISTINCT keyword).

Similarly, SQL/ADQL queries can be employed in the Gaia and MAST science archives to derive relevant data sets as described in the main text. For example, the sample of reliable stars input into the generation of a training set of reliable stars for field E0070 as described in Section~\ref{sec:reliablestars} can be generated in the Gaia archive search facility as follows:
\begin{verbatim}
SELECT ra, dec, pmra, pmdec
FROM gaiadr3.gaia_source
WHERE ruwe < 1.4 AND 
      1 = CONTAINS( POINT('ICRS', ra, dec), 
		  CIRCLE('ICRS', 216.08370, 29.34827, 4.25))
\end{verbatim}
Note that the cut on \verb+ruwe+ was not employed in generating training sets of spurious objects, as described in Section~\ref{sec:spurious}. The ICRS field centre quoted in this query was derived by precession of the B1950.0 equatorial coordinates (215.535, 29.574) appearing in the output from the first query to the SSA database above in columns \verb+(raPnt, decPnt)+. The circular area selection, achieved via the `cone--search' geometric function, specifies a radius of 4.25 degrees around the field centre to enclose the entire plate area.

\vfill 

\noindent
For the purpose of open access, the author has applied a Creative Commons Attribution (CC BY) licence to any Author Accepted Manuscript version arising from this submission. This is a pre-copyedited, author--produced PDF of an article accepted for publication in RAS Techniques and Instruments following peer review.


\bsp	
\label{lastpage}
\end{document}